\documentclass[11pt]{article}
\usepackage{epsfig}
\begin{document}
\title{A relativistic description of hadronic decays of the meson $\pi_{1}$}
\author{Nikodem Pop\l awski\footnote{\copyright\,2004 American Institute of Physics 
}\\ \\ \small{\it Nuclear Theory Center, Indiana University, Bloomington, IN 47405}\\ \scriptsize{\it This article is based on the author's Ph.D. thesis currently being written under the supervision of}\\ \scriptsize{\it Prof. Adam Szczepaniak.}}
\date{}

\maketitle
\begin{abstract}
The subject of this work is analysis of hadronic decays of exotic meson $\pi_{1}$ in a fully relativistic formalism, and comparison with the nonrelativistic results. The relativistic spin wave functions of mesons and hybrids are constructed based on unitary representations of the Lorentz group. The radial wave functions are obtained from phenomenological considerations of the mass operator. We find that decay channels $\pi_{1}\rightarrow\pi b_{1}$ and $\pi_{1}\rightarrow\pi f_{1}$ are favored, in agreement with results obtained using other models, thus indicating some model independence of the $S+P$ selection rules. We will also report on effects of meson final state interactions in exotic channels.
\end{abstract}

\section{Introduction}

In a region around 2 GeV a new form of hadronic matter is expected to exist in which the gluonic degrees of freedom are excited. In mesons these can result in resonances with exotic $J^{PC}$ quantum numbers. The adiabatic potential calculations show $\pi_{1}$ ($1^{-+}$) as the lowest energy excited gluonic configuration \cite{1}. The present models of hybrid decays (for instance \cite{2,3}) are nonrelativistic and therefore one should investigate corrections arising from fully relativistic treatment. The case of $\pi_{1}$ is of a special interest also because its evidence has been reported by the E852 collaboration and new experimental searches are planned for JLab and GSI.

\section{Relativistic spin wave function for mesons and hybrids}

For a system of non-interacting particles the spin wave function is constructed as an element of an irreducible representation of the Poincare group. We will assume $m_{u}=m_{d}=m$. In the rest frame of a quark-antiquark pair
\[ l^{\mu}_{q}=(E(m_{q},{\bf q}),{\bf q}),\,\,l^{\mu}_{\bar{q}}=(E(m_{\bar{q}},-{\bf q}),-{\bf q}), \]
the normalized spin-1 wave function ($J^{PC}=1^{--}$) is given by the Clebsch-Gordan coefficients and can be written in terms of Dirac spinors as
\begin{equation}
\Psi^{\lambda_{q\bar{q}}}_{q\bar{q}}({\bf q},{\bf l}_{q\bar{q}}=0,\sigma_{q},\sigma_{\bar{q}})=\frac{1}{\sqrt{2}m_{q\bar{q}}}\bar{u}({\bf q},\sigma_{q})\Bigl[\gamma^{i}-\frac{2q^{i}}{m_{q\bar{q}}+2m}\Bigr]v(-{\bf q},\sigma_{\bar{q}})\epsilon^{i}(\lambda_{q\bar{q}}),
\end{equation}
where $m_{q\bar{q}}$ is the invariant mass and $\epsilon^{i}(\lambda_{q\bar{q}})$ are polarization vectors corresponding to spin $1$ quantized along the z-axis. The wave function of a $q\bar{q}$ system moving with a total momentum ${\bf l}_{q\bar{q}}={\bf l}_{q}+{\bf l}_{\bar{q}}$ is given by
\begin{equation}
\Psi^{\lambda_{q\bar{q}}}_{q\bar{q}}({\bf q},{\bf l}_{q\bar{q}},\lambda_{q},\lambda_{\bar{q}})=\sum_{\sigma_{q},\sigma_{\bar{q}}}\Psi^{\lambda_{q\bar{q}}}_{q\bar{q}}({\bf q},{\bf l}_{\bar{q}}=0,\sigma_{q},\sigma_{\bar{q}})D^{\ast(1/2)}_{\lambda_{q}\sigma_{q}}({\bf q},{\bf l}_{q\bar{q}})D^{(1/2)}_{\lambda_{\bar{q}}\sigma_{\bar{q}}}(-{\bf q},{\bf l}_{q\bar{q}}),
\end{equation}
where the Wigner rotation matrix
\[ D^{(1/2)}_{\lambda\lambda'}({\bf q},{\bf P})=\Bigl[\frac{(E(m,{\bf q})+m)(E(M,{\bf P})+M)+{\bf P}\cdot{\bf q}+i{\bf \sigma}\cdot({\bf P}\times{\bf q})}{\sqrt{2(E(m,{\bf q})+m)(E(M,{\bf P})+M)(E(m,{\bf q})E(M,{\bf P})+{\bf P}\cdot{\bf q}+mM)}}\Bigr]_{\lambda\lambda'} \] 
corresponds to a boost with ${\bf \beta}\gamma={\bf P}/M$. One can show
\begin{equation}
\Psi^{\lambda_{q\bar{q}}}_{q\bar{q}}({\bf q},{\bf l}_{q\bar{q}},\lambda_{q},\lambda_{\bar{q}})=-\frac{1}{\sqrt{2}m_{q\bar{q}}}\bar{u}({\bf l}_{q},\lambda_{q})\Bigl[\gamma^{\mu}-\frac{l^{\mu}_{q}-l^{\mu}_{\bar{q}}}{m_{q\bar{q}}+2m}\Bigr]v({\bf l}_{\bar{q}},\lambda_{\bar{q}})\epsilon_{\mu}({\bf l}_{q\bar{q}},\lambda_{q\bar{q}}), 
\end{equation}
where $\epsilon^{\mu}({\bf l}_{q\bar{q}},\lambda_{q\bar{q}})$ are obtained from $(0,\epsilon^{i}(\lambda_{q\bar{q}}))$ through a boost with ${\bf \beta}\gamma={\bf l}_{q\bar{q}}/m_{q\bar{q}}$. Similarly the normalized wave function for the spin-0 quark-antiquark pair ($J^{PC}=0^{-+}$) is given by
\begin{equation}
\Psi_{q\bar{q}}({\bf q},{\bf l}_{q\bar{q}},\lambda_{q},\lambda_{\bar{q}})=\Psi_{q\bar{q}}({\bf l}_{q},{\bf l}_{\bar{q}},\lambda_{q},\lambda_{\bar{q}})=\frac{1}{\sqrt{2}m_{q\bar{q}}}\bar{u}({\bf l}_{q},\lambda_{q})\gamma^{5}v({\bf l}_{\bar{q}},\lambda_{\bar{q}}). 
\end{equation}

By coupling (3) or (4) for ${\bf l}_{q\bar{q}}=0$ with one unit of the orbital angular momentum $L=1$ and then making a boost (2), one obtains respectively the spin wave function for a quark-antiquark pair with quantum numbers $J^{PC}=1^{+-}$
\begin{equation}
\Psi^{\lambda_{q\bar{q}}}_{q\bar{q}}({\bf l}_{q},{\bf l}_{\bar{q}},\lambda_{q},\lambda_{\bar{q}})=\frac{1}{\sqrt{2}m_{q\bar{q}}({\bf l}_{q},{\bf l}_{\bar{q}})}\bar{u}({\bf l}_{q},\lambda_{q})\gamma^{5}v({\bf l}_{\bar{q}},\lambda_{\bar{q}})Y_{1\lambda_{q\bar{q}}}(\bar{{\bf q}}),
\end{equation}
or $0^{++}$, $1^{++}$ and $2^{++}$
\[ \Psi^{\lambda_{q\bar{q}}}_{q\bar{q}}({\bf l}_{q},{\bf l}_{\bar{q}},\lambda_{q},\lambda_{\bar{q}})=-\sum_{\lambda,l}\frac{1}{\sqrt{2}m_{q\bar{q}}}\bar{u}({\bf l}_{q},\lambda_{q})\Bigl[\gamma^{\mu}-\frac{l^{\mu}_{q}-l^{\mu}_{\bar{q}}}{m_{q\bar{q}}+2m}\Bigr]v({\bf l}_{\bar{q}},\lambda_{\bar{q}})\epsilon_{\mu}({\bf l}_{q\bar{q}},\lambda) \]
\begin{equation}
\cdot Y_{1l}(\bar{{\bf q}})<1,\lambda;1,l|J,\lambda_{q\bar{q}}>,
\end{equation}
with ${\bf q}=\Lambda({\bf l}_{q\bar{q}}\rightarrow0){\bf l}_{q}$.
In order to construct meson spin wave functions for higher orbital angular momenta $L$ one need only to replace $Y_{1l}$ with $Y_{Ll}$.

In the rest frame of the 3-body system corresponding to a $q\bar{q}$ pair with momentum $-{\bf Q}$ and transverse gluon with momentum ${\bf Q}$, the total spin wave function of the hybrid is obtained by coupling the $q\bar{q}$ spin-1 wave function (3) and the gluon wave function ($J^{PC}=1^{--}$) to a total spin $S=0,1,2$ and $J^{PC}=0^{++},1^{++},2^{++}$ states respectively, and then with one unit of the orbital angular momentum to the exotic state $1^{-+}$:
\[ \Psi^{\lambda_{ex}}_{q\bar{q}g(S)}(\lambda_{q},\lambda_{\bar{q}},\lambda_{g})=\sum_{\lambda_{q\bar{q}},\sigma=\pm1,M,l}\Psi^{\lambda_{q\bar{q}}}_{q\bar{q}}({\bf q},{\bf l}_{q\bar{q}}=-{\bf Q},\lambda_{q},\lambda_{\bar{q}})<1,\lambda_{q\bar{q}};1,\sigma|S,M> \]
\[ \cdot D^{(1)}_{\sigma\lambda_{g}}(\bar{{\bf Q}})Y_{1l}(\bar{{\bf Q}})<S,M;1,l|1,\lambda_{ex}>. \]
The spin-1 Wigner rotation matrix $D^{(1)}$ relates the gluon helicity $\sigma$ to its spin $\lambda_{g}$ quantized along the z-axis. The corresponding normalized wave functions are then given by:
\[ \Psi^{\lambda_{ex}}_{q\bar{q}g(S=0)}=\sqrt{\frac{3}{8\pi}}\sum_{\lambda_{q\bar{q}}}\Psi^{\lambda_{q\bar{q}}}_{q\bar{q}}({\bf q},-{\bf Q},\lambda_{q},\lambda_{\bar{q}})[{\bf \epsilon}^{\ast}(\lambda_{q\bar{q}})\cdot{\bf \epsilon}_{c}^{\ast}({\bf Q},\lambda_{g})][\bar{{\bf Q}}\cdot{\bf \epsilon}(\lambda_{ex})], \]
\[ \Psi^{\lambda_{ex}}_{q\bar{q}g(S=1)}=\sqrt{\frac{3}{8\pi}}\sum_{\lambda_{q\bar{q}}}\Psi^{\lambda_{q\bar{q}}}_{q\bar{q}}({\bf q},-{\bf Q},\lambda_{q},\lambda_{\bar{q}})[{\bf \epsilon}^{\ast}(\lambda_{q\bar{q}})\times{\bf \epsilon}_{c}^{\ast}({\bf Q},\lambda_{g})]\cdot[\bar{{\bf Q}}\times{\bf \epsilon}(\lambda_{ex})], \]
\begin{equation}
\Psi^{\lambda_{ex}}_{q\bar{q}g(S=2)}=\sqrt{\frac{27}{104\pi}}\sum_{\lambda_{q\bar{q}}}\Psi^{\lambda_{q\bar{q}}}_{q\bar{q}}(-{\bf Q},\lambda_{q},\lambda_{\bar{q}})\,\bar{{\bf Q}}\cdot[{\bf \epsilon}^{\ast}(\lambda_{q\bar{q}})\otimes{\bf \epsilon}_{c}^{\ast}({\bf Q},\lambda_{g})]\cdot{\bf \epsilon}(\lambda_{ex}), 
\end{equation}
where $\bar{{\bf Q}}={\bf Q}/|{\bf Q}|,\,\,(A\otimes B)_{ij}=2A_{(i}B_{j)}-\frac{2}{3}\delta_{ij}({\bf A}\cdot{\bf B})$ and $\epsilon^{i}_{c}({\bf Q},\lambda_{g})=\epsilon^{j}(\lambda_{g})(\delta^{ij}-\bar{Q}^{i}\bar{Q}^{j})$.

\section{Meson and hybrid states}

The $\pi\,(I=1)$ and $\eta\,(I=0)$ states ($J^{PC}=0^{-+}$) are constructed in terms of annihilation and creation operators:
\[ |M({\bf P},I,I_{3},\lambda)>\,\,=\sum_{all\,\,\lambda,c,f}\int\frac{d^{3}{\bf p}_{q}}{(2\pi)^{3}2E(m,{\bf p}_{q})}\frac{d^{3}{\bf p}_{\bar{q}}}{(2\pi)^{3}2E(m,{\bf p}_{\bar{q}})}2(E(m,{\bf p}_{q})+E(m,{\bf p}_{\bar{q}})) \]
\[ \cdot(2\pi)^{3}\delta^{3}({\bf p}_{q}+{\bf p}_{\bar{q}}-{\bf P})\frac{\delta_{c_{q}c_{\bar{q}}}}{\sqrt{3}}\frac{F(I,I_{3})_{f_{q}f_{\bar{q}}}}{\sqrt{2}}\Psi_{q\bar{q}}({\bf p}_{q},{\bf p}_{\bar{q}},\lambda_{q},\lambda_{\bar{q}})\delta_{\lambda 0}\frac{1}{N(P)} \]
\begin{equation}
\cdot\psi_{L}(\frac{m_{q\bar{q}}({\bf p}_{q},{\bf p}_{\bar{q}})}{\mu})\,b^{\dag}_{{\bf p}_{q}\lambda_{q}f_{q}c_{q}}d^{\dag}_{{\bf p}_{\bar{q}}\lambda_{\bar{q}}f_{\bar{q}}c_{\bar{q}}}|0>.
\end{equation}
In the above $\Psi_{q\bar{q}}$ is the spin-0 wave function (4), $I$ denotes isospin and $I_{3}$ is its third component, $c$ is color and $f$ is flavor. $\psi_{L}$ represents the orbital wave function resulting from the interaction between quarks that leads to a bound state (meson). Such a function depends (due to covariance) only on the invariant mass $m_{q\bar{q}}({\bf p}_{q},{\bf p}_{\bar{q}})$ \cite{4}. Normalization constants are denoted by $N$ (with $P=|{\bf P}|$) and $\mu$'s are free parameters, being scalar functions of meson quantum numbers. Finally, $F(I,I_{3})$ is 2x2 isospin matrix ($f=1$ for $u$ and $f=2$ for $d$)
\[ F(0,0)=1,\,\,\,F(1,I_{3})=\sigma^{i}\epsilon^{i}(I_{3}). \]
The flavor structure of the $\eta$ state (as well as of all meson states with isospin $0$) was chosen as a linear combination $\frac{1}{\sqrt{2}}(|u\bar{u}>+\,|d\bar{d}>)$. Strictly speaking, those states are rather linear combinations $\,a|u\bar{u}>+\,b|d\bar{d}>+\,c|s\bar{s}>$. But $|s\bar{s}>$ does not contribute to the amplitude of the decay of $\pi_{1}$ and therefore may be neglected in calculations, provided this amplitude is multiplied by a factor $\sqrt{1-|c|^{2}}$.

Similarly the $\rho\,(I=1)$ and $\omega,\,\phi\,(I=0)$ states ($J^{PC}=1^{--}$) are given by (8), but instead of $\Psi_{q\bar{q}}^{\lambda}\delta_{\lambda 0}$ one must use (3). The $b_{1}\,(I=1)$ and $h_{1}\,(I=0)$ states ($J^{PC}=1^{+-}$) contain the wave function (5) with ${\bf q}=\Lambda({\bf P}\rightarrow0){\bf p}_{q}$. Finally, the $a\,(I=1)$ and $f\,(I=0)$ states ($J^{PC}=0,1,2^{++}$) correspond to (6).

The hybrid state in its rest frame is given by
\[ |\pi_{1}(I_{3},\lambda_{ex})>\,\,=\sum_{all\,\,\lambda,c,f}\frac{1}{N_{ex}}\int\frac{d^{3}{\bf p}_{q}}{(2\pi)^{3}2E(m,{\bf p}_{q})}\frac{d^{3}{\bf p}_{\bar{q}}}{(2\pi)^{3}2E(m,{\bf p}_{\bar{q}})}\frac{d^{3}{\bf Q}}{(2\pi)^{3}2E(m_{g},{\bf Q})} \]
\[ \cdot(2\pi)^{3}2(E(m,{\bf p}_{q})+E(m,{\bf p}_{\bar{q}})+E(m_{g},{\bf Q}))\delta^{3}({\bf p}_{q}+{\bf p}_{\bar{q}}+{\bf Q})\frac{\lambda^{c_{g}}_{c_{q}c_{\bar{q}}}}{2}\frac{\sigma^{i}_{f_{q}f_{\bar{q}}}\epsilon^{i}(I_{3})}{\sqrt{2}} \]
\begin{equation}
\cdot\Psi^{\lambda_{ex}}_{q\bar{q}g}({\bf p}_{q},{\bf p}_{\bar{q}},\lambda_{q},\lambda_{\bar{q}},\lambda_{g})\psi_{L}'(\frac{m_{q\bar{q}}({\bf p}_{q},{\bf p}_{\bar{q}})}{\mu_{ex}},\frac{m_{q\bar{q}g}({\bf p}_{q},{\bf p}_{\bar{q}},{\bf Q})}{\mu_{ex'}})\,b^{\dag}_{{\bf p}_{q}\lambda_{q}f_{q}c_{q}}d^{\dag}_{{\bf p}_{\bar{q}}\lambda_{\bar{q}}f_{\bar{q}}c_{\bar{q}}}a^{\dag}_{{\bf Q}\lambda_{g}c_g}|0>,
\end{equation} 
where the spin wave function $\Psi_{q\bar{q}g}$ was given in (7) for $S=0,1,2$ and the orbital wave function $\psi_{L}'$ depends only on $m_{q\bar{q}}$ and the invariant mass of the three-body system. Here $m_{g}$ denotes the effective mass of the gluon coming from its interaction with virtual partcles, and $\lambda^{c_{g}}_{c_{q}c_{\bar{q}}}$ are the Gell-Mann matrices. Constants $N$ are fixed by normalization ($m_{M}$ is mass of meson)
\[ <{\bf P},\lambda,I_{3}|{\bf P}',\lambda',I'_{3}>\,\,=(2\pi)^{3}2E(m_{M},{\bf P})\delta^{3}({\bf P}-{\bf P}')\delta_{\lambda\lambda'}\delta_{I_{3}I_{3}'}. \]

\section{Decays of $\pi_{1}$ and nonrelativistic limit}

We will assume that a transverse gluon in $\pi_{1}$ creates a quark-antiquark pair and therefore the hybrid decays into two mesons. The Hamiltonian of this process in the Coulomb gauge is given by
\[ H=\sum_{all\,\,c,f}\int d^{3}{\bf x}\,\bar{\psi}_{c_{1}f_{1}}({\bf x})(g{\bf \gamma}\cdot{\bf A}^{c_{g}}({\bf x}))\psi_{c_{2}f_{2}}({\bf x})\delta_{f_{1}f_{2}}\frac{1}{2}\lambda^{c_{g}}_{c_{1}c_{2}}, \]
where
\[ \psi_{cf}({\bf x})=\sum_{\lambda}\int\frac{d^{3}{\bf k}}{(2\pi)^{3}2E(m,{\bf k})}[u({\bf k},\lambda)b_{{\bf k}\lambda cf}+v(-{\bf k},\lambda)d^{\dag}_{-{\bf k}\lambda cf}]e^{i{\bf k}\cdot{\bf x}} \]
and
\begin{equation}
{\bf A}^{c_{g}}({\bf x})=\sum_{\lambda}\int\frac{d^{3}{\bf k}}{(2\pi)^{3}2E(m_{g},{\bf k})}[{\bf \epsilon}_{c}({\bf k},\lambda)a^{c_{g}}_{{\bf k}\lambda}+{\bf \epsilon}_{c}^{\ast}(-{\bf k},\lambda)a^{\dag c_{g}}_{-{\bf k}\lambda}]e^{i{\bf k}\cdot{\bf x}}.
\end{equation}
Here $g$ is the strong coupling constant. The matrix element $<M_{1}({\bf P}_{1})|<M_{2}({\bf P}_{2})|H|\pi_{1}>$ ($M$ stands for meson) will be a sum of two terms because pairs $b,b^{\dag}$ and $d,d^{\dag}$ appear twice and one can show they are equal. If $\mu_{\eta}=\mu_{\pi}$ then $A_{\pi\eta}=0$ and the hybrid will not decay into $\pi$ and $\eta$. The same occurs for $\rho+\omega$. Neither can it decay into two pions because of a relative minus sign from isospin that makes both terms cancel out. However, $\mu_{b_{1}}=\mu_{\pi}$ does not imply $A_{\pi b_{1}}=0$ because the orbital wave functions of these mesons are different. 

Since $\pi$ and $\eta$ have the same quantum numbers (except isospin) and therefore $\mu_{\pi}$ and $\mu_{\eta}$ should be almost equal (not exactly because $SU(3)_{f}$ is only an approximate symmetry and there is a contribution of $s\bar{s}$ in $\eta$), out of two channels $\pi\eta$, $\pi b_{1}$ the latter will be favored. However, free parameters $\mu$ need not to be close to each other for two mesons with different radial quantum numbers, making corresponding channels significant.

A nonrelativistic limit is obtained by ignoring Wigner rotation and using nonrelativistic phase space. For $m$ large compared to $\mu$'s and $P$ this limit should be approached by relativistic results. In this case for decays of $\pi_{1}$ into $\pi$ and the S-meson $S_{q\bar{q}g}\not=1$, whereas for those into $\pi$ and the P-meson $S_{q\bar{q}g}\not=0$. The difference in amplitudes coming from the spin wave function can be clearly seen, assuming $m_{\eta}=m_{\rho},\,\mu_{\eta}=\mu_{\rho}$ and $\,m_{b_{1}}=m_{f_{1}}=m_{f_{2}},\,\mu_{b_{1}}=\mu_{f_{1}}=\mu_{f_{2}}$ (the second condition is satisfied with a good approximation by masses of particles):
\[ \Gamma_{\pi\rho}=\frac{1}{2}\Gamma_{\pi\eta},\,\,\,\Gamma_{\pi f_{1}}=\frac{1}{8}\Gamma_{\pi b_{1}}, \]
where $A_{\pi\eta}$, $A_{\pi b_{1}}$ are taken for $S=0$ and $A_{\pi\rho}$, $A_{\pi f_{1,2}}$ for $S=1$. Relations $A_{\pi\eta}$ vers. $A_{\pi b_{1}}$ and $A_{\pi\rho}$ vers. $A_{\pi f_{1,2}}$ depend on the orbital angular momentum wave functions $\psi_{L}$ and $\psi'_{L}$. If $\mu_{\rho}=\mu_{\pi}$ then in a nonrelativistic limit $\pi_{1}$ would not decay into $\pi+\rho$. Therefore the width rate for this process is expected to be much smaller than that of $\pi b_{1}$.

\section{Orbital wave function}

The orbital angular momentum wave function for a meson or a hybrid depends on the potential between quark and antiquark or for a $q\bar{q}g$ system. An explicit form of such a potential is not known exactly and such a function must be modeled. Because of the Lorentz invariance it may depend on momenta only through the invariant mass of particles. Moreover, it must tend to zero for large momenta fast enough to make the amplitude convergent. The most natural choice is the exponential function 
\[ \psi_{L}(m_{q\bar{q}}({\bf p}_{q},{\bf p}_{\bar{q}})/\mu)=e^{-m^{2}_{q\bar{q}}({\bf p}_{q},{\bf p}_{\bar{q}})/8\mu^{2}} \]
for a meson, and 
\[ \psi'_{L}(m_{q\bar{q}}({\bf p}_{q},{\bf p}_{\bar{q}})/\mu_{ex},m_{q\bar{q}g}({\bf p}_{q},{\bf p}_{\bar{q}},{\bf Q})/\mu'_{ex})=e^{-m^{2}_{q\bar{q}}({\bf p}_{q},{\bf p}_{\bar{q}})/8\mu^{2}_{ex}}e^{-m^{2}_{q\bar{q}g}({\bf p}_{q},{\bf p}_{\bar{q}},{\bf Q})/8\mu'^{2}_{ex}} \]
for a hybrid. The integrals for the decay amplitude are not elementary and must be computed numerically. In a nonrelativistic limit, however, they can be expressed in terms of the error function. 

The free parameters of the presented model are $m$, $m_{g}$, $\mu$'s and $g$. The pion form factor constants $f_{\pi}$ and $F_{\pi}$ (whose behaviour is experimentally known) defined by
\begin{equation}
<0|A^{\mu,i}({\bf 0})|\pi^{k}({\bf p})>=f_{\pi}p^{\mu}\delta_{ik},\,\,\,<\pi^{i}({\bf p}')|V^{\mu,j}({\bf 0})|\pi^{k}({\bf p})>=F_{\pi}(p^{\mu}+p'^{\mu})i\epsilon_{ijk},
\end{equation}
allow us to fit $m$ and $\mu_{\pi}$ (with the $\pi$ state given by (8)). The axial and the vector currents are defined by
\begin{equation}
A^{\mu,i}({\bf 0})=\bar{\psi}_{cf}({\bf 0})\gamma^{\mu}\gamma_{5}\frac{\sigma^{i}}{2}\psi_{cf}({\bf 0}),\,\,\,V^{\mu,j}({\bf 0})=\bar{\psi}_{cf}({\bf 0})\gamma^{\mu}\frac{\sigma^{j}}{2}\psi_{cf}({\bf 0}),
\end{equation}
with $\psi_{cf}({\bf x})$ given in (10). By virtue of the Lorentz invariance $f_{\pi}$ is a constant, whereas $F_{\pi}$ is a function of $Q^{2}=-({\bf p}-{\bf p}')^{2}$. Impossibility of finding the generators $H$ and $M_{0i}$ of the Poincare group in presence of interaction together with normalization of states violate the Lorentz covariance between spatial and time components but do not break a rotational symmetry. The resulting form factors will depend on the frame of reference.

\section{Results and summary}

Taking $m_{\pi}=(3*140+770)/4=612MeV$ (in normalization) and $m=306MeV$ gives $\mu_{\pi}=220MeV$. Assuming also $g^{2}=10$, $m_{g}=500MeV$ \cite{5}, $m_{ex}=1.6GeV$ and equality of all parameters $\mu$ leads to the following values: $\Gamma_{\pi b_{1}}=$150MeV, $\Gamma_{\pi f_{1}}=$20MeV, $\Gamma_{\pi\rho}=$3MeV.
In the nonrelativistic limit one obtains respectively: 230, 31 and 0 MeV.

Two important conclusions come from this work. Firstly, numerical results show that relativistic corrections arising from a Wigner rotation are significant. Therefore, models with no Wigner rotation are either nonrelativistic or inconsistent. These corrections decrease in general (for reasonable values of $m$) the width rates for $\pi_{1}$ or make them different from zero if they vanished for the NR case. Secondly, the $\pi_{1}$ prefers to decay into two mesons, one of which has no orbital angular momentum and the other has $L=1$ ($S+P$ selection rule). This is in agreement with other models, for instance \cite{6}. Calculations involving decay rates of $\pi_{1}$ into strange mesons and final state interactions ($b_{1}\rightarrow\pi+\omega,\,\,\omega+\pi\rightarrow\rho$) are in preparation.

\end{document}